\documentclass[authorversion,sigconf,nonacm]{acmart}
\AtBeginDocument{%
  }

\usepackage{xcolor}

\begin{document}

\title{Explainable Information Retrieval in the Audit Domain}

\author{Alexander Frummet}
\affiliation{%
  \institution{dab: GmbH}
  \city{Deggendorf}
  \country{Germany}
}
\email{alexander.frummet@dab-gmbh.de}

\author{Emanuel Slany}
\affiliation{%
  \institution{dab: GmbH}
  \city{Deggendorf}
  \country{Germany}
}
\email{emanuel.slany@dab-gmbh.de}

\author{Jonas Amling}
\affiliation{%
  \institution{dab: GmbH}
  \city{Deggendorf}
  \country{Germany}
}
\email{jonas.amling@dab-gmbh.de}

\author{Moritz Lang}
\affiliation{%
  \institution{dab: GmbH}
  \city{Deggendorf}
  \country{Germany}
}
\email{moritz.lang@dab-gmbh.de}

\author{Stephan Scheele}
\affiliation{%
  \institution{OTH Regensburg}
  \city{Regensburg}
  \country{Germany}
}
\email{stephan.scheele@oth-regensburg.de}

\renewcommand{\shortauthors}{Frummet et al.}

\begin{abstract}
  Conversational agents such as Microsoft Copilot and Google Gemini assist users with complex search tasks but often generate misleading or fabricated references. This undermines trust, particularly in high-stakes domains such as medicine and finance. Explainable information retrieval (XIR) aims to address this by making search results more transparent and interpretable. While most XIR research is domain-agnostic, this paper focuses on auditing -- a critical yet underexplored area. We argue that XIR systems can support auditors in completing their complex task. We outline key challenges and future research directions to advance XIR in this domain.
\end{abstract}





\maketitle

\section{Introduction}
Recently, agents like Microsoft Copilot and Google Gemini have gained popularity by combining the capabilities of Large Language Models (LLMs) with search systems to support users in completing complex search tasks~\citep{white2023tasks,suri2024theuse}. To promote trust, these systems often include references to the sources their answers are based on~\citep{white2023tasks}. However, these references are frequently fabricated or misleading~\citep{Walters2023}, which undermines the trustworthiness of such agents~\citep{azzopardi2024report,zhai2024large,xiong2024when}.
This issue becomes particularly critical in high-stakes domains such as medicine~\citep{fernandezpichel2025evaluating} and business~\citep{zhao2024revolutionizingfinancellmsoverview}, where decisions based on inaccurate or misleading information can have serious consequences. For instance, if doctors rely on incorrect agent responses to epilepsy~\citep{kim2024assessing} or cancer-related~\citep{rydzewski2024comparative} queries, they may make harmful treatment decisions~\citep{holzinger2017medicaldomain}. Similarly, businesses that base their strategies on misleading information risk significant financial losses~\citep{eigner2024determinants}. Therefore, building trust in these systems is essential. To this end, researchers argue that the search results underlying the agents' answers must be made more transparent, interpretable, and ultimately explainable~\citep{azzopardi2024report,culpepper2018swirl}.

To achieve this, research on explainable AI (XAI) and explainable information retrieval (XIR) has introduced methods to help users interpret model decisions~\citep{anand2022explainable,verberne2018explainable}.
XAI techniques aim to provide insights into the behaviour of models~\citep{Schwalbe2023}. 
Local XAI techniques explain model decisions given specific input instances to users.
Famous examples include feature attribution methods such as LIME~\citep{Ribeiro2016} or SHAP~\citep{Lundberg2017}
or counterfactual explanations~\citep{Wachter2017,White2020}.
Similarly, XIR research focuses on making search systems more transparent and interpretable~\citep{anand2022explainable,singh2019exs}. This includes explaining how the system interprets user queries~\citep{anand2023query,wang2024quids,zhang2020query}, developing evaluation metrics to assess explainability~\citep{chen2024evaluating,guo2023towards}, and providing explanations for search results~\citep{chen2023towards,lajewska2024explainability,leonhardt2023extractive,llordes2023explain}. Methods for search result explanations range from explaining the ranking of retrieved evidences~\citep{chen2023towards,leonhardt2023extractive,polley2021exdocs,yu2022towards} to providing additional information on the retrieved sources to enhance transparency and build trust in the system~\citep{lajewska2024explainability}. Explanations can take the form of visualisations or textual descriptions ~\citep{abujabal2017quint,lajewska2024explainability} offering users varying levels of detail~\citep{abujabal2017quint,nishida2019answering,yang2018hotpotqa}.

Most existing research focuses on domain-agnostic use cases. Search system explainability in specific domains, however, remains underexplored. While some studies have examined explainable IR in healthcare and medical applications~\citep{holzinger2017medicaldomain}, other critical fields, such as finance, law, and audit, have received little attention. Studying domain-specific applications is crucial, as each field has unique challenges and requirements. For instance, information needs in domains like cooking~\citep{frummet2022cook,frummet2019detecting} are expressed differently to those in web~\citep{broder2002taxonomy,jansen2009using} and mobile search~\citep{church2014large}. 

This paper focuses on the audit domain, a critical yet underexplored area in the IR community. Audits involve examining financial transactions for inconsistencies and fraud, requiring auditors to gather data from multiple sources, such as SAP tables and financial reports, and compile accurate reports detailing their findings.
Auditing is complex, as it involves analysing large datasets and making informed decisions. Since every major company undergoes an annual audit, an explainable search system that supports auditors’ needs could provide significant benefits. By offering transparent and interpretable results, such a system could improve efficiency, enhance accuracy, and aid in fraud detection, ultimately saving costs and ensuring financial integrity.

Therefore, we believe that explainable IR systems can assist auditors in generating comprehensive audit reports. The system should enable auditors to express complex information needs while providing transparent and interpretable retrieval results. This would help auditors assess the relevance of key financial documents and tables in their investigations.
To advance research in this domain, we propose future research directions (see Section~\ref{sec:research-directions}) and outline key challenges in XIR research for the audit context (see Section~\ref{sec:challenges}).

\section{Research Directions}
\label{sec:research-directions}
In this section, we outline key research directions to advance research on XIR in the audit domain. These include understanding auditor information needs (see Section~\ref{sec:understanding-info-needs}), designing suitable explanations (see Section~\ref{sec:explanations-audit-ir}), and developing interpretable knowledge representations (see Section~\ref{sec:knowledge-representation}).

\subsection{Information Needs in the Audit Domain}
\label{sec:understanding-info-needs}
Understanding user information needs is a critical first step in designing effective information retrieval (IR) systems, as it influences how search systems should be structured~\citep{frummet2024cooking, kelly2015development, Wildemuth2014Untangling}. For instance,~\citet{frummet-elsweiler-2024-decoding} found that users in the cooking domain tend to prefer concise answers to simple, fact-based questions (e.g., ``How much sugar do I need?'') and more detailed responses for complex, competence-oriented queries (e.g., ``How do I chop an onion?'').
Similarly, the audit domain likely involves a broad spectrum of information needs. It seems plausible that information needs in this domain could range from simple fact-checking (e.g., ``What was the invoice amount for transaction X?'') to complex procedural or investigative queries (e.g., ``How was this cost justified and approved across departments?''). The nature of the information need can influence how explanations should be presented -- whether as a direct answer, a chain of events, or links to supporting documents. Future research needs to investigate in detail the type of information needs that can occur and how these are best supported through appropriate explanations.

\subsection{Explanations in Audit-Related IR}
\label{sec:explanations-audit-ir}
Once information needs are understood, the next step is to design explanations that effectively support them. Prior work has explored various explanation formats, including visualisations~\citep{abujabal2017quint,polley2021exdocs}, detailed descriptions~\citep{polley2021exdocs,wang2024quids} (e.g., knowledge graphs showing relationships between relevant documents) and textual highlights (e.g., key sentences in relevant documents)~\citep{nishida2019answering}. In the audit context, the preferred explanation format may vary depending on the task.
For example, when verifying an expense claim, auditors might prefer a short textual summary linking the claim to relevant policy documents and approval records. In contrast, when investigating a complex revenue recognition issue, a visual explanation showing dependencies across contracts, revenue entries, and timeline events could be more useful. Understanding these preferences is key to building IR systems that enhance auditor efficiency and trust.

\subsection{Representing Audit-Related Knowledge}
\label{sec:knowledge-representation}
From a system perspective, knowledge representation plays a central role in supporting explainability. Research suggests that knowledge graphs are well-suited for this, as their structure inherently conveys relationships between entities~\citep{christmann2023explainable,chen2023towards,jia2021complex}.
In the audit domain, many elements are interdependent. For instance, a revenue entry in the general ledger might depend on multiple underlying sales contracts, which in turn relate to delivery records and approval workflows. Representing these relationships in a knowledge graph allows for intuitive navigation and explanation -- for example, showing how a revenue figure is supported by a chain of contractual and operational evidence.

\section{Challenges}
\label{sec:challenges}
In this section, we highlight the challenges of conducting XIR research in the audit domain, which relate to gathering realistic information needs (Section~\ref{sec:realistic-info-needs}), addressing data availability issues (Section~\ref{sec:data-security}), and determining appropriate ground truth explanations (Section~\ref{sec:gt-explanations}).

\subsection{Getting Realistic Information Needs}
\label{sec:realistic-info-needs}
As highlighted in Section~\ref{sec:understanding-info-needs}, information needs are typically captured through search queries and categorised using information need taxonomies. In open domains like web search or cooking, collecting such queries is relatively straightforward through crowdsourcing platforms such as Prolific or MTurk~\citep{frummet2024qooka,choi2022wot}, logging interactions with search interfaces~\citep{jansen2006searchlog,msmarco2018}, or conducting in-situ observational studies~\citep{frummet2022cook}. However, in specialised and high-stakes domains such as medicine and auditing, this becomes more difficult. These fields are closed, expert-driven environments where access is limited, making it challenging to log real user interactions or rely on crowdsourcing methods in a large scale. 

\subsection{Barriers to Using Real-World Audit Data}
\label{sec:data-security}
As discussed in Section~\ref{sec:knowledge-representation}, developing effective retrieval algorithms requires careful consideration of how audit-relevant documents are structured. However, financial data is highly confidential, and companies operate under strict regulatory and compliance standards. As a result, they may be unwilling or not allowed to share real-world data for XIR research. This makes it challenging to design and evaluate explainable retrieval systems using realistic datasets. Releasing such data for reproducibility is equally difficult and may be only possible in anonymised form.
Therefore, it is crucial to ensure that anonymised datasets still retain enough fidelity to reflect real-world scenarios. For example, anonymised employee expense reports -- preserving categories like travel, meals, or lodging, along with approval status and justifications -- can be used to simulate retrieval tasks, such as finding similar past expenses.

\subsection{Ground Truth Explanations}
\label{sec:gt-explanations}
As emphasised in Section~\ref{sec:explanations-audit-ir}, designing helpful explanations is an important research direction in this domain. A major challenge lies in providing appropriate ground truth explanations for evaluation~\citep{Schwalbe2023,Teso2019,Slany2023,Rosenfeld2021}. Research in XAI has shown that the quality of explanations is context-dependent~\citep{Heidrich2023} and that different user groups might prefer different explanation modalities~\citep{Finzel2021}. For example, a junior auditor might need detailed textual evidence pointing to inconsistencies in SAP exports, while a manager might prefer a visual overview linking anomalies across reports. These varying needs, combined with the opacity of existing systems, highlight the importance of developing and evaluating multi-modal, post-hoc explanation methods tailored to audit tasks.
\bibliographystyle{styles/ACM-Reference-Format}
\bibliography{references}

\end{document}